\documentclass[twocolumn,prd,aps,groupedaddress,showpacs,nofootinbib]{revtex4}

\usepackage{amsmath}
\usepackage{amssymb}
\usepackage{latexsym}
\usepackage{graphicx}
\usepackage{hyperref}
\usepackage{mathrsfs}
\usepackage{color}
\usepackage{bm}
\usepackage[caption=false]{subfig}

\begin{document}

\title{Estimating cosmological parameters by the simulated data of gravitational waves from the Einstein Telescope}
\author{Rong-Gen Cai}
\email{cairg@itp.ac.cn}

\author{Tao Yang}
\email{yangtao@itp.ac.cn}

\affiliation{CAS Key Laboratory of Theoretical Physics, Institute of Theoretical Physics,
Chinese Academy of Sciences, P.O. Box 2735, Beijing 100190, China and\\
School of Physical Sciences, University of Chinese Academy of Sciences, No. 19A Yuquan Road, Beijing 100049, China}


\begin{abstract}
We investigate the constraint ability of the gravitational wave (GW) as the standard siren  on the cosmological parameters by using the third-generation gravitational wave detector: the Einstein Telescope. The binary merger of a neutron with either a neutron or black hole is hypothesized to be the progenitor of a short and intense burst of $\gamma$ rays, some fraction of those binary mergers could be detected both through electromagnetic  radiation and  gravitational waves. Thus we can determine both the luminosity distance and redshift of the source separately. We simulate the luminosity distances and redshift measurements from 100 to 1000 GW events. We use two different algorithms to constrain the cosmological parameters. For the Hubble constant $H_0$ and dark matter density parameter $\Omega_m$, we adopt the Markov chain Monte Carlo approach. We find that with about 500-600 GW events we can constrain the Hubble constant with an accuracy comparable to \textit{Planck} temperature data and \textit{Planck} lensing combined results, while for the dark matter density, GWs alone seem not able to provide the constraints as good as for the Hubble constant; the sensitivity of 1000 GW events is a little lower than that of \textit{Planck} data. It should require more than 1000 events to match the \textit{Planck} sensitivity. Yet, for analyzing the more complex dynamical property of dark energy, i.e., the equation of state $w$, we adopt a new powerful nonparametric method: the Gaussian process. We can reconstruct $w$ directly from the observational luminosity distance at every redshift. In the low redshift region, we find that about 700 GW events can give the constraints of $w(z)$ comparable to the constraints of a constant $w$ by \textit{Planck} data with type Ia supernovae. Those results show that  GWs as the standard sirens to probe the cosmological parameters can provide an independent and complementary alternative to current experiments.
\end{abstract}
\maketitle

\section{Introduction \label{sec:intro}}
The last two decades have witnessed  rapid technological advances in observational cosmology. Various observations such as type-Ia supernova (SNIa)~\cite{Riess:1998cb,Perlmutter:1998np,Suzuki:2011hu,Betoule:2014frx}, the temperature and polarization anisotropy power spectrum of the cosmic microwave background (CMB) radiation~\cite{Hinshaw:2012aka,Ade:2015xua}, and weak gravitational lensing~\cite{Kilbinger:2008gk} have all indicated a Universe with an accelerated expansion. A possible explanation of this cosmic acceleration is provided by the introduction of a fluid with negative pressure called dark energy. A simple dark energy candidate, i.e., the cosmological constant $\Lambda$ whose equation of state $w =-1$ together with the cold dark matter (CDM) (called the $\Lambda$CDM model) is now called the standard model, fits the current observational data sets very well. However, some problems still exist and should be solved. For example, there is a strong tension between the value of the Hubble constant derived from the CMB~\cite{Ade:2015xua} and the value from local measurements~\cite{Riess:2011yx}. Moreover, understanding the physical property of dark energy, for example, whether it is dynamical ($w\neq-1$) or not, is one of the main challenges of modern cosmology

Though we can now measure the cosmological parameters precisely from various observations and we can further improve the capabilities of those observational methods in the future, we should note that all of the measurements are through electromagnetic (EM) radiations. In 1986, Schutz showed that it is possible to determine the Hubble constant from gravitational wave (GW) observations, by using the fact that  GWs from the binary systems encode the absolute distance information~\cite{Schutz:1986gp}. Thus the inspiraling and merging compact binaries consisting of neutron stars (NSs) and  black holes (BHs), can be considered as standard candles, or \textit{standard sirens}. The name siren is due to the fact that the GW detectors are
omnidirectional and detect coherently the phase of the wave, which  makes them in many ways more like microphones for sound than like conventional telescopes.
From the GW signal, we can measure the luminosity distance $d_L$ directly, without the need of the cosmic distance ladder: standard sirens are \textit{self-calibrating}. Assuming other techniques are available to obtain the redshift  of a GW event, for example, we can measure the redshift through the identification of an accompanying EM signal; we can get the $d_L-z$ (luminosity distance-redshift) relation. Thus we can use the GW as an alternative source to constrain the expansion history of the Universe and the cosmological parameters and it can also be a cross-check to the EM measurements.

On 11 February 2016, the Laser Interferometer Gravitational Wave Observatory (LIGO) collaboration reported the first direct detection of the gravitational wave source GW150914~\cite{Abbott:2016blz}. This  indicates that  the era of GW astronomy and the multimessage cosmology is coming. In the last decades, several papers have studied the possibility of the GW as the standard siren~\cite{Holz:2005df,Dalal:2006qt,MacLeod:2007jd,Sathyaprakash:2009xt,Nissanke:2009kt,Zhao:2010sz,Taylor:2011fs,DelPozzo:2011yh,DelPozzo:2015bna}.
Especially in~\cite{DelPozzo:2015bna}, the authors used the simulated GW signals alone to constrain cosmological parameters even regardless of the EM counterparts.
Inspired by~\cite{Sathyaprakash:2009xt,Zhao:2010sz,Li:2013lza}, we estimate the  constraint ability of cosmological parameters by the simulated data of GWs using the Einstein Telescope (ET). The ET is  a third-generation  ground-based detector of GWs~\cite{ET}. As proposed by the design document, it consists of three colocated underground detectors, each with 10 km arm and  with a $60^\circ$ opening angle. The ET is envisaged to be ten times more sensitive in amplitude than the advanced ground-based detectors, covering the frequency range of $1-10^4$ Hz. We explore how accurately it might be possible to measure the cosmological parameters such as the Hubble constant $H_0$, dark matter density parameter $\Omega_m$, and the dark energy equation of state $w$.

In Ref.~\cite{Sathyaprakash:2009xt}, with 1000 binary neutron star coalescences, the authors used the Levenberg-Marquardt algorithm to constrain the parameters $\Omega_{\Lambda}$, $\Omega_m$ and $w$. If $w$ is the only unknown parameter, it can be measured to an accuracy with $1-\sigma$ errors of $1.4\%$ with weak lensing.
Yet. in Ref.~\cite{Zhao:2010sz}, the authors used the Fisher matrix approach also with 1000 GW events and combined the {\it Planck} CMB prior to give constraints of $\Delta w_0=0.079$ and $\Delta w_a=0.261$, which are close to the detection ability of the SNAP type-Ia Supernovae project. Here $w_0$ and $w_a$ are two parameters in the equation of state of dark energy
in the Chevallier-Polarski-Linder (CPL) parametrization.
In this paper, we take two different algorithms to constrain the cosmological parameters: the Markov chain Monte Carlo (MCMC) and the Gaussian process (GP).
For simplicity, we assume a flat universe since the spatial curvature is constrained to be very close to 0 with $|\Omega_K|<0.005$~\cite{Ade:2015xua}. For the Hubble constant and the density parameter of dark matter, we adopt the MCMC method~\cite{Lewis:2002ah} and see if it can also give similar constraints as the Levenberg-Marquardt algorithm and the Fisher matrix approach used by Refs.~\cite{Sathyaprakash:2009xt} and~\cite{Zhao:2010sz}. Especially, we want to see how many GW events we can achieve with an accuracy comparable to the most recent \textit{Planck} results~\cite{Ade:2015xua}.
To study  the dynamics of the dark energy (i.e., the evolution of the equation of state), some papers like~\cite{Zhao:2010sz} define the best pivot redshift $z_p$. Thus it can constrain $w$ better since at $z_p$ the equation of state has the minimal error. In this paper, we adopt a nonparametric method, i.e., the Gaussian process~\cite{Seikel:2012uu} to reconstruct and constrain the equation of state according to Eq.~(\ref{equa:w}). We can choose the redshift region with the best performance of the reconstruction. Comparing with the method that defines the best pivot redshift, we need not parametrize the equation of state of dark energy and can directly choose the redshift where $w$ has the minimal error.

The outline of the paper is as follows. In Sec.~\ref{sec:SGRB}, we introduce the basics of using  GWs as standard sirens in the potential ET observation. In Sec.~\ref{sec:param_esti}, we use MCMC and GP methods to constrain the cosmological parameters and to see how many GW events we can achieve with an accuracy comparable to the \textit{Planck} results. We give discussions and conclusions in Sec.~\ref{sec:discussion}.

\section{ Gravitational wave as standard siren \label{sec:SGRB}}

\subsection{The model setup}

For a  Friedmann-Robertson-Walker universe, the line element reads
\begin{equation}
d{s^2} = -d{t^2} + {a^2}(t)\left[\frac{{d{r^2}}}{{1 - K{r^2}}} + {r^2}(d{\theta ^2} + {\sin ^2}\theta d{\phi ^2})\right],
\label{equa:ds}
\end{equation}
where $t$ is the cosmic time, $a(t)$ is the scale factor whose evolution depends on the matter and energy contents of the Universe, and $K = +1,-1,0$ corresponds to a closed, open, and flat universe, respectively. We use the units in which $G=c=1$ throughout this paper. Then the luminosity distance can be written as
\begin{equation}
{d_L} =\begin{cases}
\frac{{(1 + z)}}{{{H_0}\sqrt {\Omega_K}}}\sinh (\sqrt {\Omega_K} \int_0^z {\frac{{d\tilde z}}{{E(\tilde z)}})} & \Omega_K > 0 \\
\frac{{(1 + z)}}{{H_0}}\int_0^z {\frac{{d\tilde z}}{{E(\tilde z)}}} & \Omega_K = 0 \\
\frac{{(1 + z)}}{{{H_0}\sqrt {\left|\Omega_K\right|}}}\sin (\sqrt {\left|\Omega_K\right|} \int_0^z {\frac{{d\tilde z}}{{E(\tilde z)}})} & \Omega_K < 0,
\end{cases}
\label{equa:dl}
\end{equation}
where $E(z)\equiv H(z)/H_0$, $\Omega_K\equiv -K/(a_0 H_0)^2$. With the  dark energy equation of state $w(z)=p(z)/\rho(z)$, the Hubble parameter $H(z)$ is given by Friedmann equation,
\begin{align}
H{(z)^2} = &~~H_0^2 \Big\{ (1 - {\Omega _m} - {\Omega _K})\exp \Big[3\int_0^z {\frac{{1 + w(\tilde z)}}{{1 + \tilde z}}} d\tilde z\Big] \nonumber\\
           &~~+{\Omega _m}{(1 + z)^3} + {\Omega _K}{(1 + z)^2}\Big\},
\label{equa:H}
\end{align}
where $\Omega_{m,K}$ are the matter and curvature density parameters today. Since we are mainly interested in the late time  evolution of the Universe, we can ignore the radiation component. Combining Eqs.~(\ref{equa:dl}) and~(\ref{equa:H}) and writing $D(z)=H_0(1+z)^{-1} d_L(z)$ as the normalized comoving distance, we find that the equation of state can be expressed as
\begin{align}
w(z) = &~~\Big[2(1 + z)(1 + {\Omega _K}{D^2})D'' - {(1 + z)^2}{\Omega _K}D{'^3} \nonumber\\
       &~~-2(1+z){\Omega_K}DD{'^2}+3(1+{\Omega_K}D^2)D'\Big] \nonumber\\
       &~~\Big/\Big[3\{ {(1 + z)^2}[{\Omega _K} + (1 + z){\Omega _m}]D{'^2} \nonumber\\
       &~~- (1 + {\Omega _K}{D^2})\} D'\Big].
\label{equa:w}
\end{align}
Here a prime denotes the derivative with respect to redshift. Thus we see that the equation of state of dark energy is directly related  to the luminosity distance at every redshift $z$ once we know $\Omega_K$ and $\Omega_m$.

For our simulation we have to choose a fiducial cosmological model. The exact values of the cosmological parameters will not be essential in our simulations, because we are just interested in the precision with which they can be measured. However, for consistency with the current experiment data \textit{Planck} 2015~\cite{Ade:2015xua}, we choose the cosmological parameters of the fiducial model as follows,
\begin{align}
h_0=0.678,~~\Omega_m=0.308,~~\Omega_K=0,~~w=-1,
\label{ini}
\end{align}
here $H_0=100h_0$km\,s$^{-1}$Mpc$^{-1}$.  Then our work is to test the precision with which we can recover those fiducial values in Eq.~(\ref{ini}) from a set of measured luminosity distance and redshift.

\subsection{The gravitational waves with short $\gamma$-ray bursts}
Unlike current observations of the standard candles such as SNIa~\cite{Riess:1998cb,Perlmutter:1998np,Suzuki:2011hu}, the chirping GW signals from inspiraling compact binary stars (NS and BH) can provide an absolute measure of the luminosity distance~\cite{Schutz:1986gp}.  The GW amplitude depends on the so-called chirp mass and the luminosity distance, and the chirp mass can be measured from the GW signal's phasing; we can extract luminosity distance from the amplitude.

Interferometers are sensitive to the relative difference between two distances, so-called \textit{strain}. In the transverse-traceless (TT) gauge, the strain $h(t)$ can be written as
\begin{equation}
h(t)=F_+(\theta, \phi, \psi)h_+(t)+F_\times(\theta, \phi, \psi)h_\times(t),
\end{equation}
where $F_{+,\times}$ are  the \textit{beam pattern functions}, $\psi$ is the polarization angle, and ($\theta, \phi$) are angles describing the location of the source in the sky, relative to the detector~\cite{Zhao:2010sz}. $h_+=h_{xx}=-h_{yy}$, $h_\times= h_{xy}=h_{yx}$, which are the only two independent components of the GW's tensor $h_{\alpha\beta}$ in the TT gauge. The corresponding antenna pattern functions of the ET are~\cite{Zhao:2010sz}
\begin{align}
F_+^{(1)}(\theta, \phi, \psi)=&~~\frac{{\sqrt 3 }}{2}[\frac{1}{2}(1 + {\cos ^2}(\theta ))\cos (2\phi )\cos (2\psi ) \nonumber\\
                              &~~- \cos (\theta )\sin (2\phi )\sin (2\psi )],\nonumber\\
F_\times^{(1)}(\theta, \phi, \psi)=&~~\frac{{\sqrt 3 }}{2}[\frac{1}{2}(1 + {\cos ^2}(\theta ))\cos (2\phi )\sin (2\psi ) \nonumber\\
                              &~~+ \cos (\theta )\sin (2\phi )\cos (2\psi )].
\label{equa:F}
\end{align}
since the three interferometers have $60^\circ$ with each other, arranged in an equilateral triangle, the two others' antenna pattern functions are $F_{+,\times}^{(2)}(\theta, \phi, \psi)=F_{+,\times}^{(1)}(\theta, \phi+2\pi/3, \psi)$ and $F_{+,\times}^{(3)}(\theta, \phi, \psi)=F_{+,\times}^{(1)}(\theta, \phi+4\pi/3, \psi)$, respectively.

Following~\cite{Zhao:2010sz}, we define a coalescing binary with component masses $m_1$ and $m_2$, $M=m_1+m_2$ as the total mass, $\eta=m_1 m_2/M^2$ as the symmetric mass ratio, the ``chirp mass'' as $\mathcal{M}_c=M \eta^{3/5}$, and the observed chirp mass $\mathcal{M}_{c,\rm obs}=(1+z)\mathcal{M}_{c,\rm phys}$. Below, $\mathcal{M}_c$ always denotes the observed chirp mass. We also ignore the spin because we mostly consider the binary NS, in which case the phase of the waveform is computed in the post-Newtonian formalism~\cite{Blanchet:2013haa} up to 3.5 PN.  Following~\cite{Zhao:2010sz,Li:2013lza}, we also apply the stationary phase approximation to compute the Fourier transform $\mathcal{H}(f)$ of the time domain waveform $h(t)$,
\begin{align}
\mathcal{H}(f)=\mathcal{A}f^{-7/6}\exp[i(2\pi ft_0-\pi/4+2\psi(f/2)-\varphi_{(2.0)})],
\label{equa:hf}
\end{align}
where the Fourier amplitude $\mathcal{A}$ is given by
\begin{align}
\mathcal{A}=&~~\frac{1}{d_L}\sqrt{F_+^2(1+\cos^2(\iota))^2+4F_\times^2\cos^2(\iota)}\nonumber\\
            &~~\times \sqrt{5\pi/96}\pi^{-7/6}\mathcal{M}_c^{5/6}.
\label{equa:A}
\end{align}
The constant $t_0$ denotes the epoch of the merger; $\iota$ is the angle of inclination of the binary's orbital angular momentum with the line of sight.
The definitions of the functions $\psi$ and $\varphi_{(2.0)}$ can be found in~\cite{Zhao:2010sz,Li:2013lza}.

Measuring the redshift associated to a GW event is one of the biggest challenges when using the GW as the standard siren. Several methods have been proposed  for this issue, such as the galaxy catalogue proposed by Schutz~\cite{Schutz:1986gp},  neutron star mass distribution~\cite{Markovic:1993cr}, and  the tidal deformation of neutron stars~\cite{Messenger:2011gi}. In this paper, we take  a more widely used method as in~\cite{Sathyaprakash:2009xt,Nissanke:2009kt,Zhao:2010sz}: through the identification of an accompanying EM signal, namely,  the electromagnetic counterpart of the GW event. The binary merger of a NS with either a NS (BNS) or BH (BHNS) is hypothesized to be the progenitor of a short and intense burst of $\gamma$ rays (SGRB)~\cite{Nakar:2007yr}. An EM counterpart like SGRB can provide the redshift information if the host galaxy of the event can be pinpointed. Moreover, SGRBs are likely to be strongly beamed phenomena, which allow one to constrain the inclination of the compact binary system, breaking the distance-inclination degeneracy. The expected rates of BNS and BHNS detections per year for the ET are about the order $10^3-10^7$~\cite{ET}. However, only a small fraction ($\sim 10^{-3}$) is expected to satisfy the constraint that the GW events can be accompanied  with the observation of a SGRB due to the narrow beaming angle. If we assume the detection rate is in the middle range  around $\mathcal{O}(10^5)$, we can expect to see $\mathcal{O}(10^2)$ events with the SGRB per year. So, the ET is likely to detect enough GW sources to perform precision cosmology study, even under the assumption of the EM signal counterpart.

\section{ Cosmological parameters estimation \label{sec:param_esti}}

In this section, we simulate many GW detections according to the predicted rates and distributions, and record the values and errors of luminosity distances and redshifts. We constrain the cosmological parameters by simulating many catalogues of BNS and BHNS systems. For the Hubble constant and the dark matter density parameter, we take the MCMC approach. As for the dark energy equation of state, we adopt a new nonparametric approach, the GP,  to reconstruct it.

\subsection{Simulate the gravitational wave detections}

Following~\cite{Li:2013lza}, the NS mass distribution is chosen to be uniform in the interval [1,2] $M_\odot$; here $M_\odot$ is the solar mass. The black hole mass is chosen to be uniform between [3,10] $M_\odot$. Note that the chirp mass of a black hole in the first detection of a GW by LIGO is found to be a higher value of about 30 $M_\odot$. In fact a lager mass can improve the signal-to-noise ratio (SNR) of the GW detection and lead to smaller errors of the distance measurements  [see Eq.~(\ref{equa:A})]. In this paper, we still assume a conservative distribution of black hole mass given by \cite{Fryer:1999ht}. The ratio between BHNS and BNS events is taken to be 0.03, as predicted for the Advanced LIGO-Virgo network~\cite{Abadie:2010px}. The redshift distribution of the sources as observed on Earth takes the form~\cite{Zhao:2010sz}
\begin{equation}
P(z)\propto \frac{4\pi d_C^2(z)R(z)}{H(z)(1+z)},
\label{equa:pz}
\end{equation}
where $d_C$ is the comoving distance, which is defined as $d_C(z)\equiv\int_0^z {1/H(z')dz'}$, and $R(z)$ describes the time evolution of the burst rate and takes the form~\cite{Schneider:2000sg,Cutler:2009qv}
\begin{equation}
R(z)=\begin{cases}
1+2z, & z\leq 1 \\
\frac{3}{4}(5-z), & 1<z<5 \\
0, & z\geq 5.
\end{cases}
\label{equa:rz}
\end{equation}
Since it is expected that SGRBs are strongly beamed~\cite{Abdo:2009zza,Nakar:2005bs,Rezzolla:2011da}, a coincident observation of the SGRB implies that the binary was orientated nearly face on, i.e., $\iota\approx0$. In fact the maximal inclination is about $\iota=20^\circ$; however, averaging the Fisher matrix over the inclination $\iota$ and the polarization $\psi$ with the constraint $\iota<20^\circ$ is approximately the same as taking $\iota=0$~\cite{Li:2013lza}. Therefore, when we simulate the GW source we can take $\iota=0$ and the Fourier amplitude $\mathcal{A}$ in Eq.~(\ref{equa:A}) will not then depend on the polarization angle $\psi$.

The performance of a GW detector is characterized by the one-side noise \textit{power spectral density} $S_h(f)$ (PSD). We take the noise PSD of the ET to be the same as in~\cite{Zhao:2010sz}. The combined (SNR) for the network of three independent interferometers is then
\begin{equation}
\rho=\sqrt{\sum\limits_{i=1}^{3}(\rho^{(i)})^2},
\label{euqa:rho}
\end{equation}
where $\rho^{(i)}=\sqrt{\left\langle \mathcal{H}^{(i)},\mathcal{H}^{(i)}\right\rangle}$; the inner product is defined as
\begin{equation}
\left\langle{a,b}\right\rangle=4\int_{f_{\rm lower}}^{f_{\rm upper}}\frac{\tilde a(f)\tilde b^\ast(f)+\tilde a^\ast(f)\tilde b(f)}{2}\frac{df}{S_h(f)},
\label{euqa:product}
\end{equation}
where $\tilde a(f)$ and $\tilde b(f)$ are the Fourier transforms of the functions $a(t)$ and $b(t)$. The upper cutoff frequency is dictated by the last stable orbit, $f_{\rm upper}=2f_{\rm LSO}$, where $f_{\rm LSO}=1/(6^{3/2}2\pi M_{\rm obs})$ is the orbit frequency at the last stable orbit, and $M_{\rm obs}=(1+z)M_{\rm phys}$ is the observed total mass~\cite{Zhao:2010sz}. We also take the lower cutoff frequency $f_{\rm lower}=1$ Hz. In line with the SNR threshold currently used at LIGO/Virgo analysis, a GW detection is claimed only when the three ET interferometers have a network SNR of $\rho_{\rm net}>8.0$. Since we ignore the spin of the BH, the BNS or BHNS systems can be characterized by nine parameters~\cite{Li:2013lza}. With the assumption of associated SGRBs, the location of the GW source can be pinpointed by observation of  its EM counterpart. Furthermore, Ref.~\cite{Regimbau:2012ir} showed that the mass parameters can be accurately inferred and do not have considerable correlations with other parameters. Thus in the amplitude Eq.~(\ref{equa:A}), we are left with the set of parameters \{$\iota, \psi, d_L$\}. Using the Fisher information matrix, we can estimate the instrumental error on the measurement of the luminosity distance. Suppose that the error on $d_L$ is uncorrelated with the errors on the remaining GW parameters; we can find that~\cite{Zhao:2010sz,Li:2013lza}
\begin{align}
\sigma_{d_L}^{\rm inst}\simeq \sqrt{\left\langle\frac{\partial \mathcal H}{\partial d_L},\frac{\partial \mathcal H}{\partial d_L}\right\rangle^{-1}}.
\end{align}
As $\mathcal H \propto d_L^{-1}$, we can get $\sigma_{d_L}^{\rm inst}\simeq d_L/\rho$, where $\rho$ is the combined SNR of the ET. Note here that though we have set $\iota\simeq0$ when we simulate the GW source, this is an ideal situation. When we estimate the practical uncertainty of the measurement of $d_L$, we should take into account the inclination. To account for the correlation between $d_L$ and $\iota$, we note that the maximal effect of the inclination on the SNR is a factor of 2 (between $\iota =0^{\circ}$ and $\iota = 90^{\circ}$). To give an estimation of the ability of constraining cosmological parameters using the GW standard siren at least, we double the estimate of the error on the luminosity distance~\cite{Li:2013lza}
\begin{align}
\sigma_{d_L}^{\rm inst}\simeq \frac{2d_L}{\rho}.
\label{sigmainst}
\end{align}
Furthermore,  the luminosity distance is also affected by an additional error $\sigma_{d_L}^{\rm lens}$ due to the weak lensing. As in~\cite{Sathyaprakash:2009xt,Zhao:2010sz}, we assume $\sigma_{d_L}^{\rm lens}/d_L=0.05z$. Thus, the total uncertainty on the measurement of $d_L$ is taken to be
\begin{align}
\sigma_{d_L}&~~=\sqrt{(\sigma_{d_L}^{\rm inst})^2+(\sigma_{d_L}^{\rm lens})^2} \nonumber\\
            &~~=\sqrt{\left(\frac{2d_L}{\rho}\right)^2+(0.05z d_L)^2}.
\label{sigmadl}
\end{align}

\begin{figure}
\includegraphics[width=0.4\textwidth]{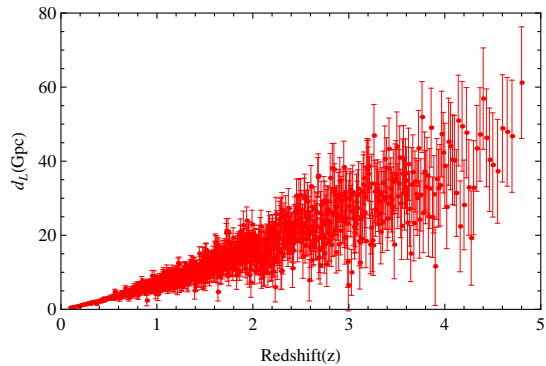}
\caption{An example catalogue with 1000 observed events of redshift, luminosity distance, and the error of the luminosity distances from the fiducial model.}
\label{fig:plotdL}
\end{figure}

As the current errors of spectroscopic redshift determination are negligible compared to the errors in the luminosity distance, we can ignore the errors of the redshift measurement by means of the EM observations. Thus, combining the fiducial model in Eq.~(\ref{ini}), the redshift distribution in Eq.~(\ref{equa:pz}), and the luminosity distance uncertainty in Eq.~(\ref{sigmadl}), we can simulate the measurements of the redshifts with the luminosity distances for the GW events of the BNS or BHNS. The basic steps are as follows. We first simulate the redshift measurements according to the redshift distribution. At every simulated redshift, we can calculate the fiducial value of the luminosity distance according to Eqs.~(\ref{equa:dl}) and~(\ref{ini}). Then we randomly sample the mass of the neutron star, the mass of the black hole, and the position angle $\theta$ in the three parameter intervals: [1,2] $M_\odot$, [3,10] $M_\odot$, and [0,$\pi$], respectively (we need not consider the other two angles $\phi$ and $\psi$ since the SNR is independent of them). Note that here we set the ratio of the possibility to detect the BHNS and BNS events $\simeq0.03$. Then we calculate the combined SNR of each set of the random sample, and confirm that it is a GW detection if $\rho_{\rm net}>8.0$. For every confirmed detection, we simulate the luminosity distance measurement $d_L^{\rm mea}$ from the fiducial value of $d_L^{\rm fid}$ and the error $\sigma_{d_L}$ in Eq.~(\ref{sigmadl}). We sample the luminosity distance measurements according to the Gaussian distribution $d_L^{\rm mea}=\mathcal{N}(d_L^{\rm fid},\sigma_{d_L})$. Thus we simulate both the redshift and the luminosity distance measurements. As we have stated before, we can expect about $10^2$ GW sources with the SGRB per year. We vary the observed number of sources from 100 up to 1000 to see that  with how many events we can constrain the cosmological parameters as precisely as the current \textit{Planck} results. An example simulating data from the fiducial model with 1000 observed events is shown in Fig.~\ref{fig:plotdL}.

\subsection{Constrain the Hubble constant and the dark matter density parameter}

To constrain $h_0$ and $\Omega_m$, we set them to be two free parameters and other parameters are fixed according to Eq.~(\ref{ini}). For a set of $N$
simulated data points,  $\chi^2$ is given by
\begin{align}
\chi^2=\sum\limits_{i=1}^{N}\left[\frac{\bar{d}_L^i-d_L(\bar{z}_i;\vec{\Omega})}{\bar{\sigma}_{d_L}^i}\right]^2
\label{equa:chi2}
\end{align}
where $\bar{z}_i$, $\bar{d}_L^i$, and $\bar{\sigma}_{d_L}^i$ are the $i$th redshift, luminosity distance, and error of luminosity distance of the simulated observational data sets. $\vec{\Omega}$ presents the set of cosmological parameters.

\begin{figure*}
\includegraphics[width=0.45\textwidth]{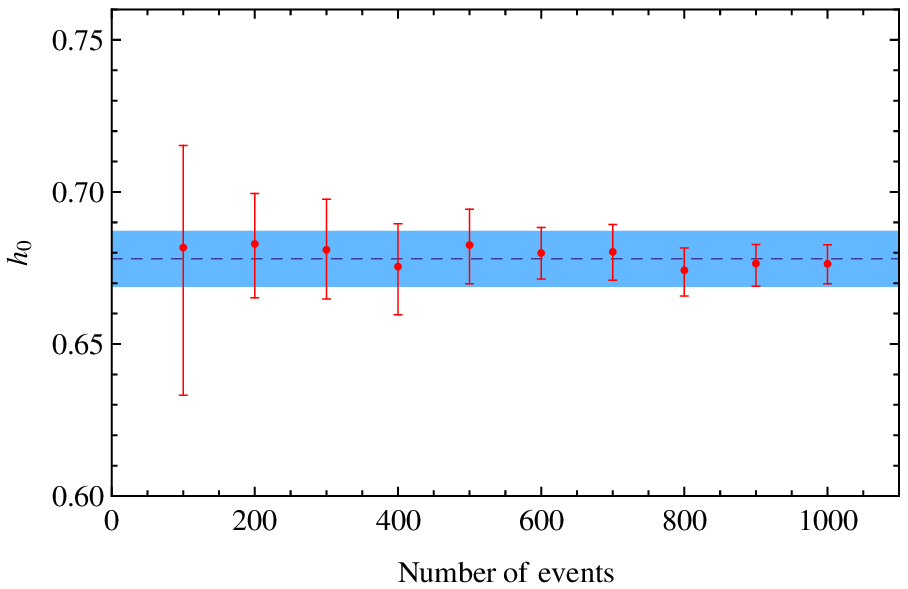}\quad
\includegraphics[width=0.45\textwidth]{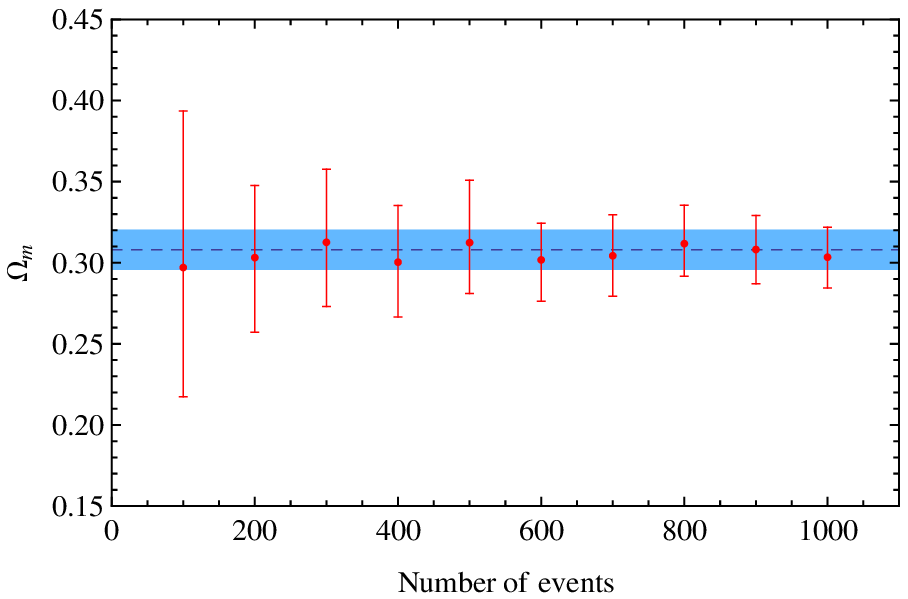}
\caption{\label{fig:homegam}Sixty-eight percent confidence level (C.L.) (red line) and the best fit (red dot) for $H_0$ (left) and $\Omega_m$ (right) for a variable number of GW events with EM counterpart. The fiducial model is shown as the dashed line. For a comparison, the blue shaded area is the $68\%$ C.L. constrained by the \textit{Planck} temperature data combined with \textit{Planck} lensing in the current \textit{Planck} 2015 results.  }
\end{figure*}

We adopt the MCMC method to find the likelihood of each parameter. As shown in Fig.~\ref{fig:homegam}, we find that with about 500-600 GW events we can constrain the Hubble constant with an accuracy comparable to \textit{Planck} temperature data and \textit{Planck} lensing combined results~\cite{Ade:2015xua}. As for the dark matter density parameter, the GW data  alone seem not able to provide a constraint as good as for the Hubble constant, the sensitivity of 1000 GW events is a little lower than that of \textit{Planck} data. It should require more than 1000 events to match the \textit{Planck} sensitivity.

\subsection{Constrain the equation of state of dark energy}

Next we turn to study the ability of the standard siren to infer the nature of dark energy. Unlike those works in Refs.~\cite{Zhao:2010sz,Li:2013lza} that define a pivot point, here we adopt a new nonparametric method, the GP, to reconstruct $w(z)$ in the whole redshift region. Here we should note that  this method has some advantages and also disadvantages. The first advantage is that we can study the nature of the dark energy in the whole redshift region. Once having reconstructed the luminosity distance, we can use it to reconstruct $w(z)$ at each redshift point as we want. Secondly, we can simply set $w(z)$ as a function of redshift $z$, and need not parametrize the equation of state like the CPL form. Thus we can constrain the equation of state more directly and model independently.  However, the GP reconstruction method has some shortcomings. Since we use only the simulated data of the luminosity distance, the errors of the reconstructed $D(z)$ heavily  depend on the quality of those simulated data. Moreover, we should reconstruct the $D(z)$'s derivatives up to second order, and then combine its derivatives to give the final reconstructed $w(z)$. We can see below that the errors of $w(z)$ become large in the high-$z$ region. On the other hand, note that the errors of the mock data of $d_L$ or $D$ are very small in the low redshift region (Fig.~\ref{fig:plotdL}). So, we just focus on the low redshift region where the reconstruction can be performed very well. Anyway, we want to use a new nonparametric method to reconstruct the equation of state from the simulated data, and check its ability of constraining $w(z)$ in the low redshift region.

The GPs allow one to reconstruct a function and its derivatives from data without assuming a parametrization for it.
We use  the GPs in Python (GaPP)~\cite{Seikel:2012uu} to derive our GP reconstruction results.
The distribution over functions  provided by the GP is suitable to describe the observed data. At each point $z$, the reconstructed function $f(z)$ is also a Gaussian distribution with a mean value and Gaussian error. The functions at different points $z$ and $\tilde{z}$ are related by a covariance function $k(z,\tilde{z})$, which only depends on a set of hyperparameters $\ell$ and $\sigma_f$. Here $\ell$ gives a measure of the coherence length of the correlation in the $x$-direction and $\sigma_f$ denotes the overall amplitude of the correlation in the $y$-direction. Both of them are optimized by the GP with the observed data set.
In contrast to actual parameters, the GP does not specify the form of the reconstructed function. Instead it characterizes the typical changes of the function.

The different choices of the covariance function may affect the reconstruction to some extent. The covariance function usually takes  the squared exponential form as~\cite{Seikel:2012uu}
\begin{equation}
k(z,\tilde{z})={\sigma_f}^2 \exp\Big(-\frac{(z-\tilde{z})}{2\ell^2}\Big).
\end{equation}
But it is not always a suitable choice. Here we take the Mat\'{e}rn ($\nu = 9/2$) covariance function
\begin{align}
k(z,\tilde z) = &~{\sigma _f}^2\exp\left( - \frac{{3\left| {z - \tilde z} \right|}}{\ell }\right) \nonumber \\
      &~~\times\Big[1 + \frac{{3\left| {z - \tilde z} \right|}}{\ell } + \frac{{27{{(z - \tilde z)}^2}}}{{7{\ell ^2}}} \nonumber \\
     &~~ + \frac{{18{{\left| {z - \tilde z} \right|}^3}}}{{7{\ell ^3}}} + \frac{{27{{(z - \tilde z)}^4}}}{{35{\ell ^4}}}\Big],
\end{align}
according to the analysis made in~\cite{Seikel:2013fda}, where they considered various assumed models and many realizations of mock data sets for a test and concluded that the Mat\'{e}rn ($\nu=9/2$) covariance function can lead to more reliable results than all others when applying GP to reconstructions using $D$ measurements.
The detailed analysis and description of the GP method can be found in~\cite{Seikel:2012uu,Seikel:2013fda}, where the authors studied the use of the GP method to reconstruct dark energy dynamics from supernovae data. Some of the GP's applications can also be found in~\cite{Nair:2013sna} and in our previous works~\cite{Cai:2015zoa,Cai:2015pia,Cai:2016vmn}.

\begin{figure*}
\subfloat[700 events]{
\includegraphics[width=0.4\textwidth]{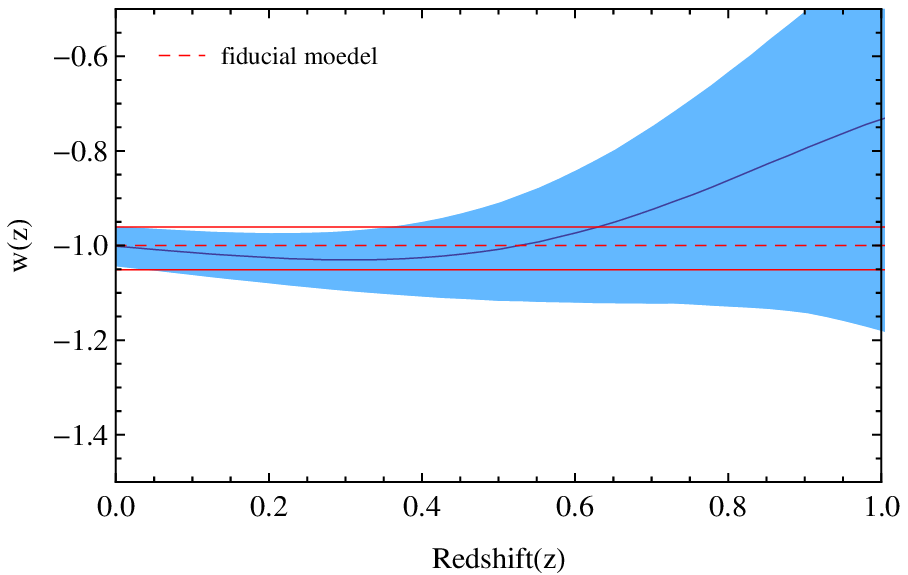}}\quad
\subfloat[800 events]{
\includegraphics[width=0.4\textwidth]{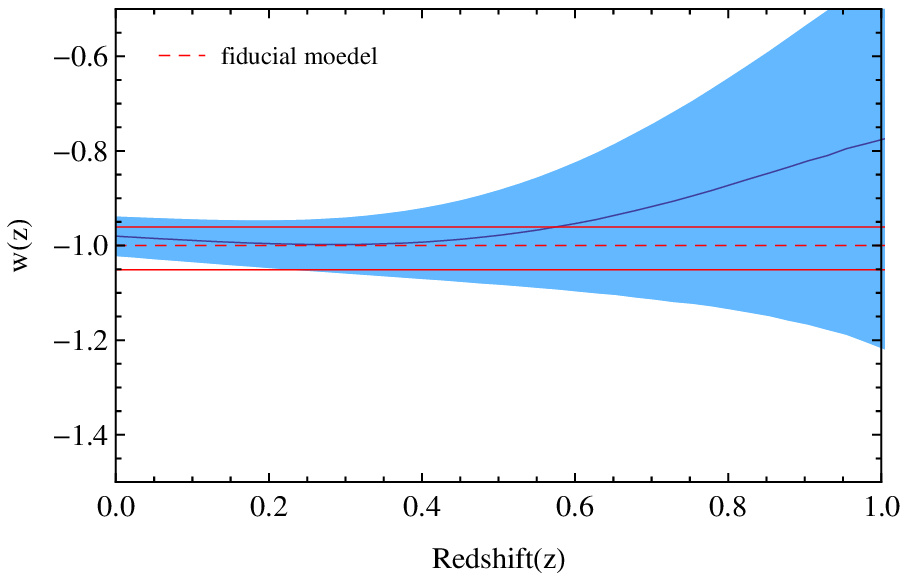}}\\
\subfloat[900 events]{
\includegraphics[width=0.4\textwidth]{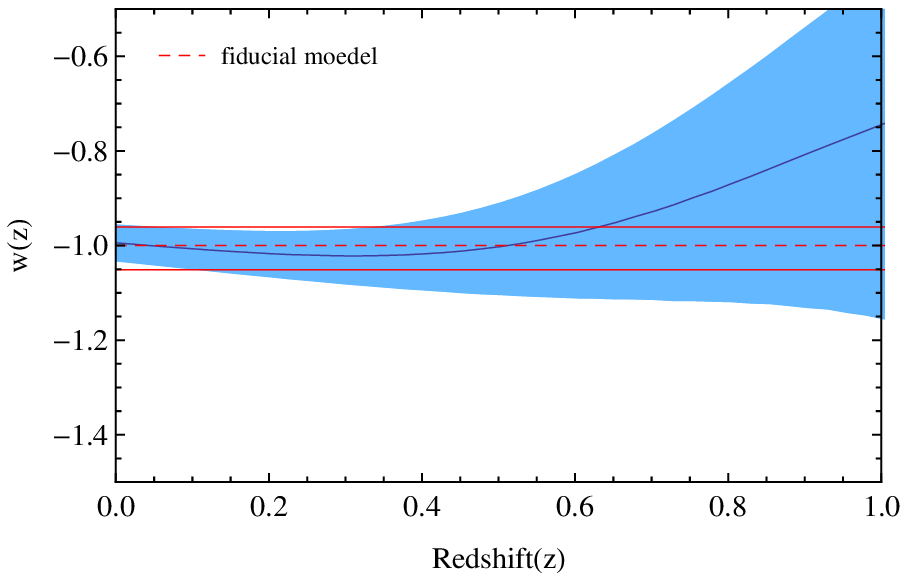}}\quad
\subfloat[1000 events]{
\includegraphics[width=0.4\textwidth]{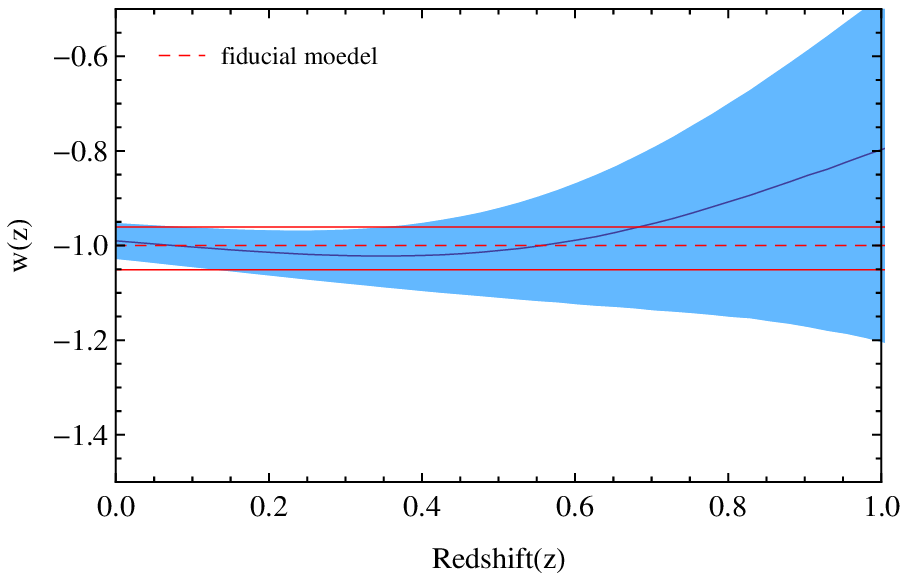}}
\caption{\label{fig:wplot}Reconstructions of $w(z)$ from the simulated data sets with a variable number of GW events. The shaded blue regions are the $68\%$ C.L. of the constraint. The fiducial value $w=-1$ is shown as the red dashed line. For a comparison, we add the two red lines that bound the $68\%$ C.L. of the constraint of the constant equation of state $w$ from the \textit{Planck} data combined with the type-Ia supernovae.}
\end{figure*}

Using Eq.~(\ref{equa:w}), we transform the reconstruction of distance $D(z)$ and its derivatives to obtain the constraint of $w(z)$  for the cases with different  numbers of GW events. We compare our results with \textit{Planck} 2015. The results are shown in Fig.~\ref{fig:wplot} for the cases with $N=700, 800, 900$, and $1000$, respectively. Since we use only the $\{z,d_L\}$ data sets to reconstruct the equation of state $w(z)$ which is dependent on $z$, while we compare it with the constant $w$ constrained by \textit{Planck} data combined with type-Ia supernovae, the reconstructed errors of $w(z)$ in our results are of course larger than the  errors in the {\it Planck} results  with a constant $w$ in the high-$z$ region. However, as shown in Fig.~\ref{fig:wplot}, we can see that 700 GW events can give the same constraint accuracy  to $w(z)$ as \textit{Planck} in the low redshift region. Thus, the GWs can be an alternative source to study the dynamics of the dark energy. It can be expected that with a better data analysis method, combining the GWs with the traditional EM data, the cosmological parameters can be constrained more precisely.

\section{Conclusions and discussions \label{sec:discussion}}

In this paper, we studied the gravitational wave as the standard siren to constrain the cosmological parameters. Gravitational waves from coalescing binaries directly encode the luminosity distance. The redshift $z$ of the sources can be determined with great accuracy through the electromagnetic counterparts. The candidate electromagnetic signal is the short $\gamma$-ray burst that is supposed to be the aftermath of the binary system with at least one neutron star. We used the ET design to study  the constraint ability on the  cosmological parameters by simulating binary systems of NS NS and NS BH that have an accompanying EM signal. We estimated the instrumental error on the luminosity distance by using the Fisher matrix approach. We also added the weak lensing errors to the instrumental error. Combing the redshift distribution with the fiducial model, we simulated the luminosity distance measurements from 100 up to 1000 GW events.

For the Hubble constant and the dark matter density parameter, we used the MCMC method to derive their likelihoods. We found that with about 500-600 GW events we can constrain the Hubble constant with an accuracy comparable to \textit{Planck} 2015 results. As for the dark matter density parameter, the GW data  alone seem not able to provide constraints as good  as for the Hubble constant; the sensitivity of 1000 GW events is a little lower than that of \textit{Planck} data. It should require more than 1000 events to match the \textit{Planck} sensitivity. With 1000 GW events, our results agree with those in Ref.~\cite{Zhao:2010sz}: $\Delta h_0\sim 5\times 10^{-3}$ and $\Delta \Omega_m\sim 0.02$. In order to study the dynamics of  dark energy, we adopted a new nonparametric method Gaussian process to reconstruct the equation of state of dark energy. For the feature of this method, we focused on the constraint in the low redshift region. We found that about 700 GW events can give constraints of $w(z)$ comparable to those of the constant $w$ by \textit{Planck} data with type-Ia supernovae. With 1000 GW events, we can constrain $w(z)$ with an error of $\Delta w(z)\sim 0.03$ in the low redshift region. Our method is more powerful and gives a better constraint than that of  Ref.~\cite{Zhao:2010sz} using ET-GW observations and the future baryon acoustic oscillation (BAO) and supernova observations of the Joint Dark Energy Mission (JDEM) project,  especially in the low redshift region. Yet, Ref.~\cite{Zhao:2010sz} uses the Fisher matrix method and CPL parametrization and gives a constraint of $\Delta w_0\sim0.077$ and $\Delta w_a=0.445$ using ET-GW observations and a comparable result using the JDEM BAO and supernova. Those results show that  the GW as a standard siren to probe the cosmological parameters can provide an independent and complementary alternative to current experiments. Moreover, the GP method is a powerful method and can give a better constraint when it is employed to study the dynamics of dark energy. It is  expected that by combining more data sets, including the EM and GW data , the GP method could play an important role in constraining cosmological parameters in the future.

Here we mention that the detection calibration and selection biases may have influences on the distance measurement. At present, the calibration constitutes about $10\%$ absolute error in advanced LIGO distance measurements. This may be reduced by the time of ET observation. Furthermore, the selection biases such as the SNR cut and the assumptions inherent in the burst rate [Eq.~(\ref{equa:rz})] should also somehow affect the accuracy of the measurements. In this paper, we follow a traditional way to handle these issues. For the burst rate bias, as suggested by Ref.~\cite{Zhao:2010sz}, we do not expect that it can produce a noticeable influence on the results.

Observing an electromagnetic counterpart is not the only method to measure the redshift associated to a GW event. The other ways to get the redshift information can be found in~\cite{Schutz:1986gp,Markovic:1993cr,Messenger:2011gi}. In addition, the spin of the BH can also help us estimate the GW's parameters.  It is expected that by combining GW data with other astronomical observations such as supernovae, and adopting a better data analysis approach, the cosmological parameters could be constrained more precisely than the current situation. We leave this for future studies.


\begin{acknowledgements}
This work is supported in part by the National Natural
Science Foundation of China Grants No. 11375247,
No. 11435006 and No. 11690022, and in part by the
Strategic Priority Research Program of the Chinese Academy of Sciences (CAS), Grants No. XDB09000000 and No. XDB 23030000, and by the key project of CAS, Grant No. QYZDJ-SSW-SYS006.

\end{acknowledgements}


\end{document}